\documentclass[twocolumn, twocolappendix]{aastex631}
\usepackage{enumitem}
\usepackage{xspace}
\usepackage{amsmath}
\usepackage{float}

\newcommand{\sw}[1]{\texttt{#1}}

\shorttitle{Selection Effects in FRB DM--$z$ Analysis}
\shortauthors{Sharma et al.}

\begin{document}
\title{Quantifying the impact of selection effects on FRB DM--$z$ relation cosmological inference}

\correspondingauthor{Kritti Sharma}
\email{kritti@caltech.edu}

\author[0000-0002-4477-3625]{Kritti Sharma}
\affiliation{Cahill Center for Astronomy and Astrophysics, MC 249-17 California Institute of Technology, Pasadena CA 91125, USA.}

\author[0000-0002-7252-5485]{Vikram Ravi}
\affiliation{Cahill Center for Astronomy and Astrophysics, MC 249-17 California Institute of Technology, Pasadena CA 91125, USA.}
\affiliation{Owens Valley Radio Observatory, California Institute of Technology, Big Pine CA 93513, USA.}

\author[0000-0002-7587-6352]{Liam Connor}
\affiliation{Center for Astrophysics | Harvard $\&$ Smithsonian, Cambridge, MA 02138-1516, USA.}

\author[0000-0001-8356-2014]{Elisabeth Krause}
\affiliation{Department of Astronomy/Steward Observatory, University of Arizona, 933 North Cherry Avenue, Tucson, AZ 85721, USA}
\affiliation{Department of Physics, University of Arizona, 1118 E Fourth Street, Tucson, AZ 85721,
USA}

\author[0000-0003-3714-2574]{Pranjal R. S.}
\affiliation{Department of Astronomy/Steward Observatory, University of Arizona, 933 North Cherry Avenue, Tucson, AZ 85721, USA}

\author[0000-0003-3312-909X]{Dhayaa Anbajagane}
\affiliation{Department of Astronomy and Astrophysics, University of Chicago, Chicago, IL 60637, USA}
\affiliation{Kavli Institute for Cosmological Physics, University of Chicago, Chicago, IL 60637, USA}

\begin{abstract}

Fast Radio Bursts (FRBs) have emerged as powerful probes of baryonic matter in the Universe, offering constraints on cosmological and feedback parameters through their extragalactic dispersion measure--redshift (DM$_\mathrm{exgal}$--$z$) relation. However, the observed FRB population is shaped by complex selection effects arising from instrument sensitivity, DM-dependent search efficiency, and FRB source population redshift-evolution. In this work, we quantify the impact of such observational and population selection effects on cosmological inference derived from the conditional distribution $p(\mathrm{DM}_{\mathrm{exgal}}|z)$. Using forward-modeled FRB population simulations, we explore progressively realistic survey scenarios incorporating redshift evolution, luminosity function, and instrument DM selection function. To enable rapid likelihood evaluations, we build a neural-network emulator for the variance in cosmic DM, $\sigma^2[\mathrm{DM}_{\mathrm{cosmic}}(z)]$, trained on $5\times10^4$ baryonification halo-model simulations, achieving $\leq4\%$ accuracy up to $z=4$. We demonstrate that while redshift and DM-dependent selection effects substantially alter the joint distribution $p(\mathrm{DM},z)$, they have a negligible impact on the conditional distribution $p(\mathrm{DM}_{\mathrm{exgal}}|z)$ for current sample sizes. The parameter biases are $\lesssim0.8\sigma$ for $10^2$ FRBs, indicating that conditional analyses are robust for present surveys. However, depending on the survey DM-dependent search efficiency, these biases may exceed $3\sigma$ for $10^4$ FRBs, thus implying that explicit modeling of selection effects will be essential for next-generation samples.

\end{abstract}

\section{Introduction} \label{sec:introduction}

Fast Radio Bursts~\citep[FRBs;][]{2007Sci...318..777L, 2013Sci...341...53T} are millisecond-duration transients of extragalactic origin~\citep{2022A&ARv..30....2P} that have emerged as powerful cosmological probes for tracing the distribution of baryons in the Universe~\citep{2014ApJ...780L..33M, 2020Natur.581..391M} at small scales upto $k\sim 10~h$\,Mpc$^{-1}$~\citep{2025ApJ...989...81S}. Their signals encode a variety of propagation effects, most notably the frequency-dependent dispersive delay caused by ionized plasma along the line of sight, quantified by the dispersion measure (DM). Since the DM reflects the integrated column density of free electrons along an FRB’s path, it offers a unique and complementary probe of the baryon distribution within and around massive halos, supplementing traditional baryon detection methods~\citep{2025arXiv250707991K, 2025arXiv250714136H, 2025arXiv250607432P, 2025arXiv250704476D}.

The modeling of the DM--redshift relation for FRBs, pioneered by \citet{2020Natur.581..391M}, established the foundation for their use as large-scale tracers of baryons. The mean of this cosmic DM relation, $\langle \mathrm{DM}_\mathrm{cosmic} (z) \rangle$ is proportional to the Hubble constant weighted by the cosmic matter density in baryons and diffuse ionized fraction of baryons in the Universe~\citep[see Equation~\ref{eqn:DMmean};][]{2020Natur.581..391M}. The sightline-to-sightline scatter, or the variance in cosmic DM for FRB sources at the same redshift depends on the degree of clustering of baryons~\citep[see Equation~\ref{eqn:DMvariance};][]{2014ApJ...780L..33M} in $\gtrsim 10^{11}~M_\odot$ halos with peak sensitivity to $\sim 10^{14}~M_\odot$ galaxy groups and clusters~\citep{2025ApJ...989...81S}. Therefore, both, the mean and the variance of the DM--redshift relation carry cosmologically and astrophysically valuable information. 

Traditionally, two main approaches have been used to model the observed distribution of FRBs in the two-dimensional DM--redshift space. The first approach models the \emph{joint distribution}, $p(\mathrm{DM}, z)$~\citep{2022MNRAS.509.4775J, 2022MNRAS.516.4862J}, while the second focuses on the \emph{conditional distribution}, $p(\mathrm{DM}|z)$~\citep{2020Natur.581..391M, 2024arXiv240916952C}. For cosmological applications, the latter approach is advantageous because it is less sensitive to the observed redshift distribution of FRB sources, which is governed by the unknown population characteristics, such as the population redshift evolution and FRB luminosity function~\citep[see Section~\ref{subsec:modeling_frb_population_and_instrument_observational_biases};][]{2022MNRAS.509.4775J, 2022MNRAS.516.4862J}. However, both approaches are susceptible to the instrument DM selection function, which depends on the specific characteristics of the FRB search, such as DM smearing across frequency channels, finite time resolution, and the maximum searched DM~\citep{2019MNRAS.487.5753C, 2022MNRAS.509.4775J, 2022MNRAS.516.4862J}. 

There is broad consensus within the community that these population and observational selection effects alter the shape of the joint probability distribution $p(\mathrm{DM}, z)$. However, some debate remains regarding the magnitude and nature of the bias they may introduce in the inferred cosmological and astrophysical feedback parameters when analyses are based on the conditional probability $p(\mathrm{DM} | z)$. Given the growing emphasis on large-sample statistics to enable use of FRBs as precision probes of baryons, especially with the advent of next generation of FRB experiments, including the Deep Synoptic Array~\citep[DSA-2000;][]{2019BAAS...51g.255H}, the Square Kilometre Array~\citep[SKA;][]{2004NewAR..48..979C}, and the Canadian Hydrogen Observatory and Radio Transient Detector~\citep[CHORD;][]{2019clrp.2020...28V}, it is timely to assess the impact of population evolution, luminosity function, and DM-dependent instrument sensitivity on cosmological inference.

In this work, we perform comprehensive mock FRB survey simulations to quantify how observational selection effects may bias the inference of cosmological and astrophysical feedback parameters in FRB analyses. We begin by reviewing the key ingredients and assumptions underlying our model FRB DM--redshift relation in Section~\ref{sec:model}. We summarize the model for our neural-network emulator of variance in cosmic DM in Section~\ref{sec:building_emulator}. In Section~\ref{sec:quantify_biases}, we present the results from a series of mock analyses, each introducing additional levels of complexity, to quantify the resulting biases, as detailed below:
\begin{enumerate}[label=(\roman*),leftmargin=*,itemsep=-0.4pt]
    \item No selection effects (see Section~\ref{subsec:no_selection_effects}),
    
    \item FRB source redshift distribution (see Section~\ref{subsec:imposing_redshift_distribution} and Figure~\ref{fig:pDMz_bias_zdist}),
    
    \item Sharp DM cut-off in FRB search (see Section~\ref{subsec:sharp_DMcut} and Figure~\ref{fig:pDMz_bias_zstar2p0_strictDMcut}), and
    
    \item Varied strength of DM cut-off in FRB search (see Section~\ref{subsec:vary_DMcut} and Figure~\ref{fig:pDMz_bias_zstar2p0_DMcut}).
\end{enumerate}
We perform analyses both, with and without modeling selection effects, across aforementioned four scenarios, allowing us to quantify the resulting parameter biases. Finally, we conclude in Section~\ref{sec:conclusion} with a summary of principal outcomes of this work. Throughout the text, we assume \citet{2020A&A...641A...6P} TT,TE,EE+lowE+lensing cosmology.

\section{The Observed Dispersion Measure - Redshift Relation} \label{sec:model}

As radio waves propagate from a source at redshift $z_\mathrm{s}$ along direction $\hat{x}$ to an observer, they are progressively dispersed by free electrons in the Milky Way, the intergalactic medium (IGM), the circumgalactic medium (CGM) of intervening halos, and the host galaxy of FRB. The observed dispersion measure (DM$_\mathrm{FRB}$) can be expressed as
\begin{equation}
\mathrm{DM}_\mathrm{FRB} (\hat{x},z_\mathrm{s}) = \mathrm{DM}_\mathrm{MW}(\hat{x}) + \mathrm{DM}_\mathrm{cosmic}(\hat{x}, z_\mathrm{s}) + \dfrac{\mathrm{DM}_\mathrm{host}}{(1+z_\mathrm{s})},
\label{eqn:DM_FRB}
\end{equation}
where the Galactic contribution ($\mathrm{DM}_\mathrm{MW}$) includes the contribution of the Milky Way interstellar medium (ISM) and halo. The NE2001/YMW16 provide models for ISM contribution with an uncertainty of $\lesssim 10$~pc\,cm$^{-3}$ for sightlines away from the Galactic plane~\citep{2002astro.ph..7156C, 2003astro.ph..1598C, 2019ascl.soft08022Y}. The Milky Way halo contribution is often estimated to be in the range 30-80~pc\,cm$^{-3}$~\citep{2019MNRAS.485..648P, 2020ApJ...888..105Y, 2023arXiv230101000R, 2023ApJ...946...58C}. The last two terms in Equation~\ref{eqn:DM_FRB} constitutes the extragalactic component ($\mathrm{DM}_\mathrm{exgal}$), which includes the contribution of cosmic gas distribution and FRB host galaxy (see Section \ref{subsec:modeling_extragalactic_dispersion_measure}).

Apart from the large-scale structure, the \textit{observed} distribution of FRBs in two-dimensional DM$_\mathrm{FRB} - z$ space is also shaped by the FRB population characteristics and instrument detection sensitivity~\citep[see Section \ref{subsec:modeling_frb_population_and_instrument_observational_biases};][]{2022MNRAS.509.4775J, 2022MNRAS.516.4862J}. The joint probability of observing an FRB with dispersion measure DM$_\mathrm{FRB}$ and redshift $z$ is
\begin{equation}
\begin{aligned}
    p(\mathrm{DM}_\mathrm{FRB}, z) = & \, p(\mathrm{DM}_\mathrm{exgal} = \mathrm{DM}_\mathrm{FRB} - \mathrm{DM}_\mathrm{MW}|z) \, \cdot \\
    & \, S(\mathrm{DM}_\mathrm{FRB}) \cdot p_\mathrm{exist}(z) \cdot p_\mathrm{detect}(z),
\end{aligned}
\label{eqn:2D_DMz_model}
\end{equation}
where 
\begin{enumerate}[label=(\roman*),leftmargin=*,itemsep=0pt]
    \item $p(\mathrm{DM}_\mathrm{exgal}|z)$ is the conditional probability that an FRB is observed with extragalactic dispersion measure $\mathrm{DM}_\mathrm{exgal}$ at redshift $z$, which probes the cosmological gas distribution through DM$_\mathrm{cosmic}$~\citep[see Equation~\ref{eqn:DM_FRB} and Section~\ref{subsec:modeling_extragalactic_dispersion_measure};][]{2014ApJ...780L..33M} and is the key observable of cosmological interest in this work~\citep{2020Natur.581..391M},
    
    \item $S(\mathrm{DM})$ is the instrument DM selection function, which is shaped by dispersion smearing in frequency channels, instrument time resolution and FRB search setup in the detection pipeline (see Section \ref{subsubsec:DM_selection_function}).
    \item $p_\mathrm{exist}(z)$ is the probability of existence of an FRB source at redshift $z$, which depends on redshift evolution of the FRB source population (see Section \ref{subsubsec:population_evolution}), and
    \item $p_\mathrm{detect}(z)$ is the probability of detecting an FRB at redshift $z$, which depends on the instrument detection fluence threshold and the FRB luminosity function (see Section \ref{subsubsec:luminosity_function_and_instrument_detection_threshold}).
\end{enumerate}
In this section, we review the key components of our model DM$_\mathrm{FRB} - z$ relation. A comprehensive discussion of this subject can be found in \citet{2022MNRAS.509.4775J, 2022MNRAS.516.4862J}.

\subsection{Extragalactic Dispersion Measure} \label{subsec:modeling_extragalactic_dispersion_measure}

The extragalactic dispersion measure $\mathrm{DM}_\mathrm{exgal}$ distribution at redshift $z$ is written as
\begin{equation}
    \begin{aligned}
        & p(\mathrm{DM}_\mathrm{exgal} | z) \\
        & = \int\limits_0^{\mathrm{DM}_\mathrm{exgal}} 
        p(\mathrm{DM}_\mathrm{cosmic}|z) \cdot p\left( \mathrm{DM}_\mathrm{host} \right) 
        \mathrm{d} \mathrm{DM}_\mathrm{cosmic},
    \end{aligned}
    \label{eqn:DMexgal}
\end{equation}
where $\mathrm{DM}_\mathrm{host} = (\mathrm{DM}_\mathrm{exgal} - \mathrm{DM}_\mathrm{cosmic}) \times (1+z)$. 

Simulations suggest that the DM$_\mathrm{host}$ distribution is well approximated by a log-normal distribution~\citep{2020ApJ...900..170Z, 2024arXiv240308611T}: 
\begin{equation}
    \begin{aligned}
        & p(\mathrm{DM}_\mathrm{host}) \\
        & = \frac{1}{\mathrm{DM}_\mathrm{host}\sigma_\mathrm{host} \sqrt{2\pi}}\mathrm{exp}\left( - \frac{(\ln \mathrm{DM}_\mathrm{host} - \mu_\mathrm{host})^2}{2\sigma_\mathrm{host}^2} \right),
    \end{aligned}
    \label{eqn:DMhost}
\end{equation}
with parameters $(\mu_\mathrm{host}, \sigma_\mathrm{host})$, such that $\langle \mathrm{DM}_\mathrm{host} \rangle = e^{\mu_\mathrm{host}}$ and $\sigma^2[\mathrm{DM}_\mathrm{host}] = e^{2\mu_\mathrm{host} + \sigma_\mathrm{host}^2}(e^{\sigma_\mathrm{host}^2}-1)$ are the mean and variance of the distribution, respectively. Throughout this work, we assume no redshift evolution of rest-frame $\mathrm{DM}_\mathrm{host}$ distribution.

The conditional probability $p(\mathrm{DM}_\mathrm{cosmic}|z)$ has been shown to be well approximated by a log-normal distribution in hydrodynamical simulations~\citep{2025ApJ...989...81S}\footnote{While we adopt a log-normal parameterization for $p(\mathrm{DM}|z)$ because its first two moments map cleanly onto observables such as the mean and variance of the cosmic dispersion measure, \citet{2025arXiv250707090K} demonstrate that $p(\mathrm{DM}_\mathrm{cosmic}|z)$ distribution in IllustrisTNG hydrosimulations departs from log-normality, particularly at its tails. It would therefore be valuable in future works to investigate whether closed-form analytical likelihoods can be derived for the more accurate functional forms proposed in \citet{2025arXiv250707090K}, enabling unbiased cosmological inference without relying on the log-normal approximation.} The mean of this distribution as a function of redshift explicitly depends on cosmological parameters,
\begin{equation}
\begin{aligned}
    \langle \mathrm{DM}_\mathrm{cosmic} (z_\mathrm{s}) \rangle & = \int\limits_0^{z_\mathrm{s}} \frac{3c \chi_\mathrm{e} \Omega_\mathrm{b0} H_0}{8 \pi G m_\mathrm{p}} \frac{f_\mathrm{d}(z) (1+z)~\mathrm{d}z}{\sqrt{\Omega_\mathrm{m0}(1+z)^3 + \Omega_{\Lambda 0}}} \\
    & = \int\limits_0^{\chi(z_s)} W_\mathrm{DM}(\chi) \mathrm{d}\chi,
\label{eqn:DMmean}
\end{aligned}
\end{equation}
where $f_\mathrm{d}(z)$ denotes the fraction of baryons in diffuse ionized gas, $\chi_\mathrm{e} = Y_\mathrm{H} + Y_\mathrm{He}/2 \approx 1 - Y_\mathrm{He}/2$ is the free electron fraction from primordial helium abundance~\citep{2020A&A...641A...6P}, $m_\mathrm{p}$ is the proton mass, $H_0$ is the Hubble constant and $\Omega_\mathrm{b0}$, $\Omega_\mathrm{m0}$ and $\Omega_{\Lambda 0}$ are the present epoch baryon, matter and dark energy densities, respectively. Here, the second equality is the equivalent integral using the comoving distance ($\chi$), where the prefactors have been folded into the DM weighting function $W_\mathrm{DM}(\chi)$.

The second moment of the $p(\mathrm{DM}_\mathrm{exgal}|z)$ distribution, the variance in $\mathrm{DM}_\mathrm{cosmic}$ at a given redshift $z$, arises from intervening collapsed structures along the line of sight~\citep{2014ApJ...780L..33M} and can be written under flat-sky and Limber approximation as (see \citealt{2023MNRAS.524.2237R, 2025ApJ...989...81S} for details):
\begin{equation}
\begin{aligned}
    \sigma^2 [\mathrm{DM}_\mathrm{cosmic}(z_\mathrm{s})] & = \int\limits_0^{\chi_\mathrm{s}} \mathrm{d}\chi W_\mathrm{DM}^2(\chi) \int\limits_0^\infty \frac{k \mathrm{d}k}{2\pi} P_\mathrm{ee} (k, z(\chi)),
\end{aligned}
\label{eqn:DMvariance}
\end{equation}
where $P_\mathrm{ee}(k,z)$ is the electron power spectrum. While we do not model the full covariance between sightlines due to the large-scale structure, such correlations will be essential to incorporate for unbiased inference with large FRB samples~\citep{2023MNRAS.524.2237R}.

\subsection{Instrument Observational Biases and FRB Population Characteristics} \label{subsec:modeling_frb_population_and_instrument_observational_biases}

In the previous section, we discussed the model for $p(\mathrm{DM}_\mathrm{exgal}|z)$, which probes the cosmological gas distribution. In this section, we discuss the FRB population characteristics and instrument selection function that shapes the two-dimensional DM distribution of observed FRBs (see Equation~\ref{eqn:2D_DMz_model}).

\subsubsection{DM Selection Function} \label{subsubsec:DM_selection_function}

The detectability of a burst with dispersion measure DM$_\mathrm{FRB}$ depends on how dispersion interacts with the FRB search setup: dispersion smearing within each frequency channel ($w_{\mathrm{DM}}$) and finite time resolution ($w_{\mathrm{samp}}$) broadens the effective pulse width $w_{\mathrm{eff}}$ as
\begin{equation}
    w_{\mathrm{eff}} = \sqrt{w_{\mathrm{int}}^2 + w_{\mathrm{scat}}^2 + w_{\mathrm{DM}}^2 + w_{\mathrm{samp}}^2},
    \label{eqn:effective_burst_width}
\end{equation}
where $w_{\mathrm{int}}$ is the intrinsic burst width and $w_{\mathrm{scat}}$ is the scattered width~\citep{2003ApJ...596.1142C}. This degrades detection efficiency of the instrument ($\eta$) to the burst as 
\begin{equation}
    \eta(w_\mathrm{eff}) \propto \left( \frac{w_\mathrm{eff}}{1~\mathrm{ms}} \right)^{-0.5}.
    \label{eqn:detection_efficiency}
\end{equation}
Furthermore, the finite DM search window limits the accessible DM range, excluding bursts outside it, and the DM search resolution, set by trial DM step size $\delta_\mathrm{DM}$, introduces residual smearing across the total bandwidth. Together, these effects reduce sensitivity at large DM and shape the DM-dependent selection function of the instrument. The DM selection function of an instrument can be measured and characterized with a robust injection recovery system~\citep{2023AJ....165..152M}. Alternatively, we consider a sigmoid function approximation to it~\citep{2018Natur.562..386S}:
\begin{equation}
    S(\mathrm{DM}_\mathrm{FRB}) \propto \frac{1}{1+\exp\left(s(\mathrm{DM}_\mathrm{FRB} - \mathrm{DM}_\mathrm{cut})\right)},
    \label{eqn:DMselectionfxn}
\end{equation}
with cut-off strength $s$ and cut-off DM of the search pipeline DM$_\mathrm{cut}$.

\subsubsection{Population Evolution} \label{subsubsec:population_evolution}

The probability of an FRB source existing at redshift $z$ is governed by the redshift evolution of volumetric source density $\Phi(z)$ as
\begin{equation}
    p_\mathrm{exist}(z) \propto \Phi(z)\,\frac{\mathrm{d}V(z)}{\mathrm{d}\Omega\,\mathrm{d}z},
    \label{eqn:p_exist}
\end{equation}
where the second term denotes the differential comoving volume element. Given that the dominant FRB formation channel is expected to trace the galaxy star-forming main sequence~\citep{2023ApJ...954...80G, 2024Natur.635...61S}, $\Phi(z)$ can be reasonably approximated as being proportional to the cosmic star-formation rate,
\begin{equation}
    \Phi(z) \propto \frac{1}{1+z}\left( \frac{\mathrm{SFR}(z)}{\mathrm{SFR}(0)} \right)^n,
    \label{eqn:rate_evolution}
\end{equation}
where the $(1+z)$ factor accounts for cosmological time dilation, and $\mathrm{SFR}(z)$ denotes the cosmic star-formation rate density~\citep{2014ARA&A..52..415M},
\begin{equation}
    \mathrm{SFR}(z) \;\propto\;
    \frac{(1+z)^{2.7}}
         {1 + \left( \dfrac{1+z}{2.9} \right)^{5.6}} .
    \label{eqn:SFR_density_evolution}
\end{equation}
The parameter $n$ introduces flexibility to account for possible deviations from a model that purely traces star-formation. In principle, the FRB source volumetric density should be written as the convolution of cosmic star-formation history with the FRB delay-time distribution~\citep{2023ApJ...950..175S, 2024ApJ...967...29L}. However, owing to limited empirical constraints on the FRB delay-time distribution, Equation~\ref{eqn:rate_evolution} provides a well-motivated first-order approximation.

\begin{table*}
    \centering
    \begin{tabular}{lllll}
        \toprule
        Parameter & Equation & Description & Prior & Fiducial \\
        \hline

        \multicolumn{3}{l}{\textbf{Baryonification/Feedback Parameters}} \\
        $\log M_c$ & \ref{eqn:gas_profile_slope} & Mass scale below which gas profile becomes shallower than NFW & U[11.0, 15.0] & 14.0 \\
        $\mu_\beta$ & \ref{eqn:gas_profile_slope} & Mass dependence of inner slope $\beta$ & - & 1 \\
        $\delta$ & \ref{eqn:gas_profile} & Outer slope of bound gas profile & - & 7 \\
        $\xi$ & \ref{eqn:gas_profile} & Outer slope of bound gas profile & - & 2.5 \\
        $\theta_\mathrm{co}$ & \ref{eqn:gas_profile} & Radius at which the inner slope of the gas density profile changes & - & 0.1 \\
        $\theta_\mathrm{ej}$ & \ref{eqn:gas_profile} & Radius at which the outer slope of the gas density profile changes & - & 3.5 \\

        \multicolumn{3}{l}{\textbf{Cosmological Parameters}} \\
        $\Omega_\mathrm{m0}$ & \ref{eqn:DMmean}, \ref{eqn:DMvariance} & Total matter density & U[0.2, 0.4] & 0.3097 \\
        $\Omega_\mathrm{b0}$  & \ref{eqn:DMmean}, \ref{eqn:DMvariance} & Baryon density & U[0.03, 0.07] & 0.0490 \\
        $H_0$ & \ref{eqn:DMmean}, \ref{eqn:DMvariance} & Hubble constant & U[50, 90] & 67.66 \\
        $\sigma_8$ & \ref{eqn:DMmean}, \ref{eqn:DMvariance} & Amplitude of density fluctuations & U[0.5, 1.0] & 0.8111 \\

        \multicolumn{3}{l}{\textbf{FRB host galaxy DM distribution}} \\
        $\mu_\mathrm{host}$ & \ref{eqn:DMhost} & Lognormal DM$_\mathrm{host}$ distribution parameter & U[4.0, 6.0] & 5.0 \\
        $\sigma_\mathrm{host}$ & \ref{eqn:DMhost} & Lognormal DM$_\mathrm{host}$ distribution parameter & U[0.2, 1.0] & 0.5 \\

        \multicolumn{3}{l}{\textbf{FRB population}} \\
        $\gamma$ & \ref{eqn:luminosity_function} & Power-law index in the FRB luminosity function & U[-0.9, 1] & -0.5 \\
        $z_\ast$ & \ref{eqn:p_detect_simplified} & Detection horizon for the upper cut-off energy of luminosity function & U[0.01, 3] & - \\
        $z_\mathrm{min}$ & \ref{eqn:p_detect_simplified} & Minimum detectable redshift & - & 0 \\
        $n$ & \ref{eqn:rate_evolution} & Redshift-evolution of volumetric FRB source density & - & 1 \\
        $\alpha$ & \ref{eqn:E_th} & FRB spectral index & - & -0.5 \\ 

        \multicolumn{3}{l}{\textbf{Instrument Sensitivity}} \\
        DM$_\mathrm{cut}$ & \ref{eqn:DMselectionfxn} & DM cut-off & U[1000, 5000] & - \\
        $s$ & \ref{eqn:DMselectionfxn} & Strength/steepness of the DM cut-off & U[-3, 0] & - \\

        \hline
    \end{tabular}
    \caption{Summary of DM$_\mathrm{FRB} - z$ model parameters. The priors for parameters varied when fitting simulated datasets and fiducial values for all parameters in our simulations are listed.}
    \label{table:parameters}
\end{table*}

\subsubsection{Luminosity Function} \label{subsubsec:luminosity_function_and_instrument_detection_threshold}

The probability of a burst at redshift $z$ occurring with energy $E$ above the detection threshold $E_\mathrm{th}(z)$ of the instrument can be written as
\begin{equation}
    p_\mathrm{detect}(z) = p(E > E_\mathrm{th}(z)) = \dfrac{\int\limits_{\mathrm{max}(E_\mathrm{min}, E_\mathrm{th}(z))}^{\infty} \Phi(E) \mathrm{d}E}{\int\limits_{E_\mathrm{min}}^{\infty} \Phi(E) \mathrm{d}E},
    \label{eqn:p_detect}
\end{equation}
where we assume a Schechter luminosity function~\citep{1976ApJ...203..297S} distribution of burst energies,
\begin{equation}
    \Phi(E) \mathrm{d}E = \Phi_\ast \left( \frac{E}{E_\ast} \right)^\gamma \exp \left( -\frac{E}{E_\ast} \right) \mathrm{d}\left( \frac{E}{E_\ast} \right),
    \label{eqn:luminosity_function}
\end{equation}
$\Phi_\ast$ is the volumetric density of event rate, $\gamma$ is the power-law index, $E_\ast$ is the upper cut-off energy and $E_\mathrm{min}$ is the minimum burst energy. The detection threshold $E_\mathrm{th}$ depends on the effective fluence threshold as
\begin{equation}
\begin{aligned}
    E_\mathrm{th}(z) & = F_\mathrm{th,eff} \Delta \nu \frac{4\pi D_L^2(z)}{(1+z)^{2+\alpha}} \\
    & = F_\mathrm{th,eff} \cdot \zeta(z)
\end{aligned}
\label{eqn:E_th}
\end{equation}
where $\Delta \nu$ is the instrument bandwidth, $\alpha$ is the FRB spectral index $F \propto \nu^\alpha$, $D_L$ is the luminosity distance and $(1+z)^2$ factor in the denominator accounts for photon energy redshifting and time dilation. The $\zeta(z)$ function in the second equality captures the redshift evolution of $E_\mathrm{th}(z)$. 

The fluence threshold $F_\mathrm{th}$ is a function of the beam sensitivity $B$ and detection efficiency $\eta$,
\begin{equation}
    F_\mathrm{th}(w_\mathrm{eff}, \Omega) = \frac{F_1}{\eta(w_\mathrm{eff}) B(\Omega)},
\label{eqn:F_th}
\end{equation}
where $F_1$ is the nominal fluence threshold calculated using the radiometer equation, referenced to a 1~ms duration burst at the beam center. The relative beam sensitivity $B$ as a function of position on sky with-respect-to the bore-sight $\Omega$, is normalized to a maximum of 1 at the beam center and the detection efficiency $\eta$ for a burst of effective width $w_\mathrm{eff}$ is defined in Equation~\ref{eqn:detection_efficiency}. Given this model, the effective fluence threshold $F_\mathrm{th,eff}$ can be written by marginalizing over the telescope beamshape and the burst width distribution $p(w_\mathrm{eff})$ as
\begin{equation}
    F_\mathrm{th,eff} = \dfrac{\int \mathrm{d}B\,\Omega(B) \int \mathrm{d}w_\mathrm{eff}\,p(w_\mathrm{eff})\,F_\mathrm{th}(w_\mathrm{eff}, \Omega)}{\int \mathrm{d}B\,\Omega(B) \int \mathrm{d}w_\mathrm{eff}\,p(w_\mathrm{eff})},
    \label{eqn:F_eff,th}
\end{equation}
where $\Omega(B)$ is the inverse beamshape. Note that we assume the effective fluence threshold is a constant quantity, marginalized over the telescope beamshape and the burst width distributions. With this constant effective fluence threshold, we can define minimum detectable redshift $z_\mathrm{min}$ as $E_\mathrm{min} = F_\mathrm{th, eff} \zeta(z_\mathrm{min})$ and detection horizon $z_\ast$ as
$E_\ast = F_\mathrm{th, eff} \zeta(z_\ast)$. Plugging everything back in Equations~\ref{eqn:p_detect} and \ref{eqn:luminosity_function}, the detection probability of an FRB at redshift $z$ can be simplified to be written as 
\begin{equation}
    p_\mathrm{detect}\!\left(z\right) =
    \frac{
        \Gamma\!\left[\gamma+1, 
        \dfrac{\zeta\left(\max\!\bigl[z_\mathrm{min}, z\bigr]\right)}{\zeta\left(z_\ast\right)}
        \right]
    }{
        \Gamma\!\left[\gamma+1, 
        \dfrac{\zeta\left(z_\mathrm{min}\right)}
        {\zeta\left(z_\ast\right)}\right]
    },
\label{eqn:p_detect_simplified}
\end{equation}
where $\Gamma(.)$ denotes the Gamma function.

\section{Emulator for Variance in Cosmic Dispersion Measure} \label{sec:building_emulator}

In this section, we present the emulator for $\sigma^2[\mathrm{DM}_\mathrm{cosmic}(z)]$ as a function of cosmological and astrophysical feedback parameters. We describe the baryonification model for gas power spectrum in Section~\ref{subsec:baryonification_power_spectrum_model}, which is used to simulate $\sigma^2[\mathrm{DM}_\mathrm{cosmic}(z)]$ variations with cosmology and feedback. We then use baryonification simulations under the halo model framework to build a neural network model that accurately emulates $\sigma^2[\mathrm{DM}_\mathrm{cosmic}(z)]$ in Section~\ref{subsec:neural_network_emulator}.

\subsection{Baryonification Power Spectrum Model} \label{subsec:baryonification_power_spectrum_model}

The extragalactic DM distribution $p(\mathrm{DM}_\mathrm{exgal} |z)$ depends on the variance in DM$_\mathrm{cosmic}$ (see Equation~\ref{eqn:DMvariance}). The feedback dependence of $\sigma^2[\mathrm{DM}_\mathrm{cosmic}(z)]$ is explicit through the $P_\mathrm{ee}(k,z)$, which we compute under halo model framework. Specifically, we calculate the gas power spectrum using baryonification gas profiles~\citep{2015JCAP...12..049S, 2019JCAP...03..020S}, as implemented in \sw{BaryonForge} software~\citep{2024OJAp....7E.108A}. The radial gas profile $\rho_\mathrm{gas}$ of a halo of mass $M$ is given by~\citep{2021JCAP...12..046G}
\begin{equation}
    \rho_\mathrm{gas}(r|M) = \dfrac{\rho_\mathrm{gas, 0}}{\left[ 1+\left( \dfrac{r}{\theta_\mathrm{co} r_\mathrm{200c}} \right)^\beta \right] \left[ 1+\left( \dfrac{r}{\theta_\mathrm{ej} r_\mathrm{200c}} \right)^\xi \right]^{\frac{\delta-\beta}{\xi}}},
    \label{eqn:gas_profile}
\end{equation}
where $\rho_\mathrm{gas, 0}$ is a normalization constant. The halo mass dependent slope of the gas profile ($\beta$) can be written as
\begin{equation}
    \beta = \dfrac{3(M/M_c)^{\mu_\beta}}{1+(M/M_c)^{\mu_\beta}},
    \label{eqn:gas_profile_slope}
\end{equation}
where $M_c$ is the halo mass below which the gas profile becomes shallower than the NFW profile and $\mu_\beta$ controls the halo mass scaling of the slope. The remaining slope parameters $\{\delta, \gamma\}$, core radius $\theta_\mathrm{co}$ and ejected radius $\theta_\mathrm{ej}$ do not have an explicit mass dependence. In our models, we only consider variations of $M_c$ in the range $10^{11} - 10^{16}~M_\odot$ and fix rest of the parameters to the fiducial values in \citet{2021JCAP...12..046G}. In future, when we have better constraining power with large FRB samples, it would be worth exploring more complex models which consider variations of other baryonification profile parameters listed in Table~\ref{table:parameters}.

\subsection{Neural Network Emulator} \label{subsec:neural_network_emulator}

We build a neural network--based emulator to model the dependence of $\sigma [\mathrm{DM}_\mathrm{cosmic}(z)]$ on cosmological and astrophysical parameters over the redshift range $z \in [0, 4]$. We train the emulator on a Latin Hypercube-sampled suite of $5\times 10^4$ simulations spanning the parameter space defined in Table~\ref{table:parameters}. The emulator input consist of four cosmological parameters $\{\Omega_\mathrm{m0}, \Omega_\mathrm{b0}, H_0, \sigma_8\}$ and one feedback parameter $\{\log M_c\}$, while the output corresponds to discretized $\sigma [\mathrm{DM}_\mathrm{cosmic}]$ in redshift $z$. 

Before training, the simulation outputs $\sigma [\mathrm{DM}_\mathrm{cosmic}(z)]$ were smoothed using a Savitzky--Golay filter with window length 15 and polynomial order 3 to suppress numerical noise while preserving the broadband features of $\sigma [\mathrm{DM}_\mathrm{cosmic}(z)]$. To reduce the high-dimensional output space, we perform Incremental Principal Component Analysis (IPCA) and retain the first four principal components with highest eigen values. This choice captures over 99\% of the total variance in each field while enabling efficient training. We perform PCA fits in batches of size 256 to accommodate large datasets. The filtered fields $X$ are then projected into PCA space as $\mathbf{T} = (\mathbf{X} - \boldsymbol{\mu}) \mathbf{W}$, 
where $\mathbf{W}$ denotes the matrix of PCA eigenvectors and $\boldsymbol{\mu}$ the mean field. The PCA coefficients $\mathbf{T}$ serve as regression targets for the neural network emulator.

The emulator employs a fully connected feed-forward neural network with two hidden layers, each containing 400 neurons and ReLU activation functions. The network architecture can be summarized as:
\begin{equation}
    \text{Input (5)} \rightarrow 2 \times \text{Dense (400, ReLU)} \rightarrow \text{Dense (4)}.
\end{equation}  
The output layer is linear, allowing the model to regress continuous PCA coefficients directly.

We train the model using Adam optimizer (learning rate $10^{-3}$) with a mean squared error (MSE) loss function. We apply early stopping with a patience of 100 epochs based on validation loss to prevent overfitting. We use training and validation split of 99\%--1\%. After training, the emulator predictions $\hat{\mathbf{T}}$ are inverse transformed through the PCA basis to reconstruct the predicted fields as $\hat{\mathbf{X}} = \hat{\mathbf{T}}\,\mathbf{W}^{\top} + \boldsymbol{\mu}$.

All preprocessing and PCA steps are implemented in \texttt{scikit-learn}, while the neural network is built and trained using \texttt{TensorFlow/Keras}. Models, PCA objects, and scalers were serialized using \texttt{HDF5} and \texttt{joblib} for downstream inference. The full training and evaluation pipeline was executed on GPU-accelerated hardware with batch sizes of 128 for neural network training and 256 for PCA transformations.

\begin{figure}
\centering
\includegraphics[width=\columnwidth]{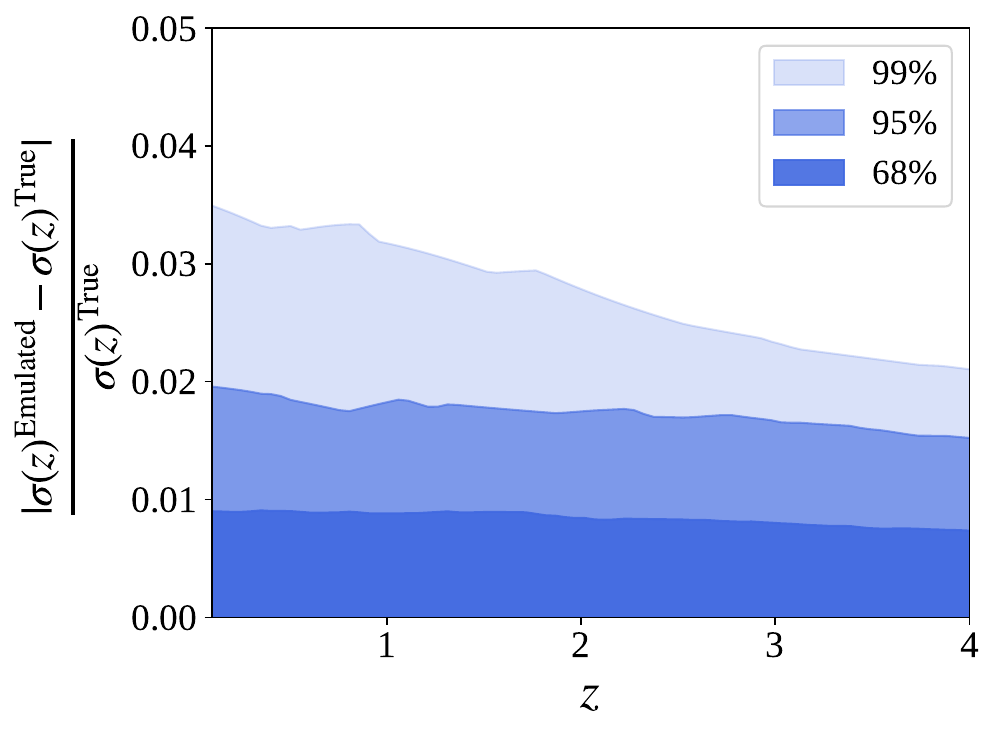}
\caption{Accuracy of the emulator for FRB dispersion measure variance, $\sigma [\mathrm{DM}_\mathrm{cosmic}]$ as a function of redshift $z$, under the halo model prescription~\citep{2024OJAp....7E.108A} with baryonification halo gas profiles~\citep{2015JCAP...12..049S, 2019JCAP...03..020S}. We show the 68\%, 95\% and 99\% upper limits on the distribution of error in emulated variance compared to the true variance. We achieve $\leq$4\% accuracy up to redshift 4 on validation dataset.}
\label{fig:emulator}
\end{figure}

We quantify the emulator performance by comparing predicted and true simulation outputs via the metric:
\begin{equation}
    R_{\sigma} = \frac{{\sigma}(z)^\mathrm{Emulated}}{\sigma(z)^\mathrm{True}} -1.
\end{equation}
We demonstrate the accuracy of our model by computing the percentile bands (16--84\% and 5--95\%) for metric $R_{\sigma}$ across all test realizations in Figure~\ref{fig:emulator}. The reconstructed fields typically agree with the true simulation outputs at $\lesssim 4$\%-level across all redshifts, demonstrating the fidelity of the emulator.

\section{Quantitative Assessment of Selection-Induced Biases} \label{sec:quantify_biases}

The ultimate goal of this exercise is to quantify the bias in cosmological inference conducted using FRBs due to FRB population evolution and instrument selection effects. In practice, this boils down to asking that given the observed distribution of FRBs in two-dimensional DM$_\mathrm{FRB} - z$ space, what is the bias in inference conducted using $p(\mathrm{DM}_\mathrm{exgal}|z)$ alone, without modeling population and instrument selection effects. To quantify the parameter biases, in this section, we begin with no selection effects scenario and in successive experiments, introduce more complexity to the model from FRB population and instrument DM selection function. The mock observed FRBs are sampled from the $p(\mathrm{DM}_\mathrm{exgal}, z)$ distribution in each successive experiment. Finally, we discuss the possible limitations of our selection effects modeling in Section~\ref{subsec:limitations}.

\subsection{No Selections Effects} \label{subsec:no_selection_effects}

\begin{figure*}[ht!]
\centering
\fbox{\includegraphics[width=\textwidth]{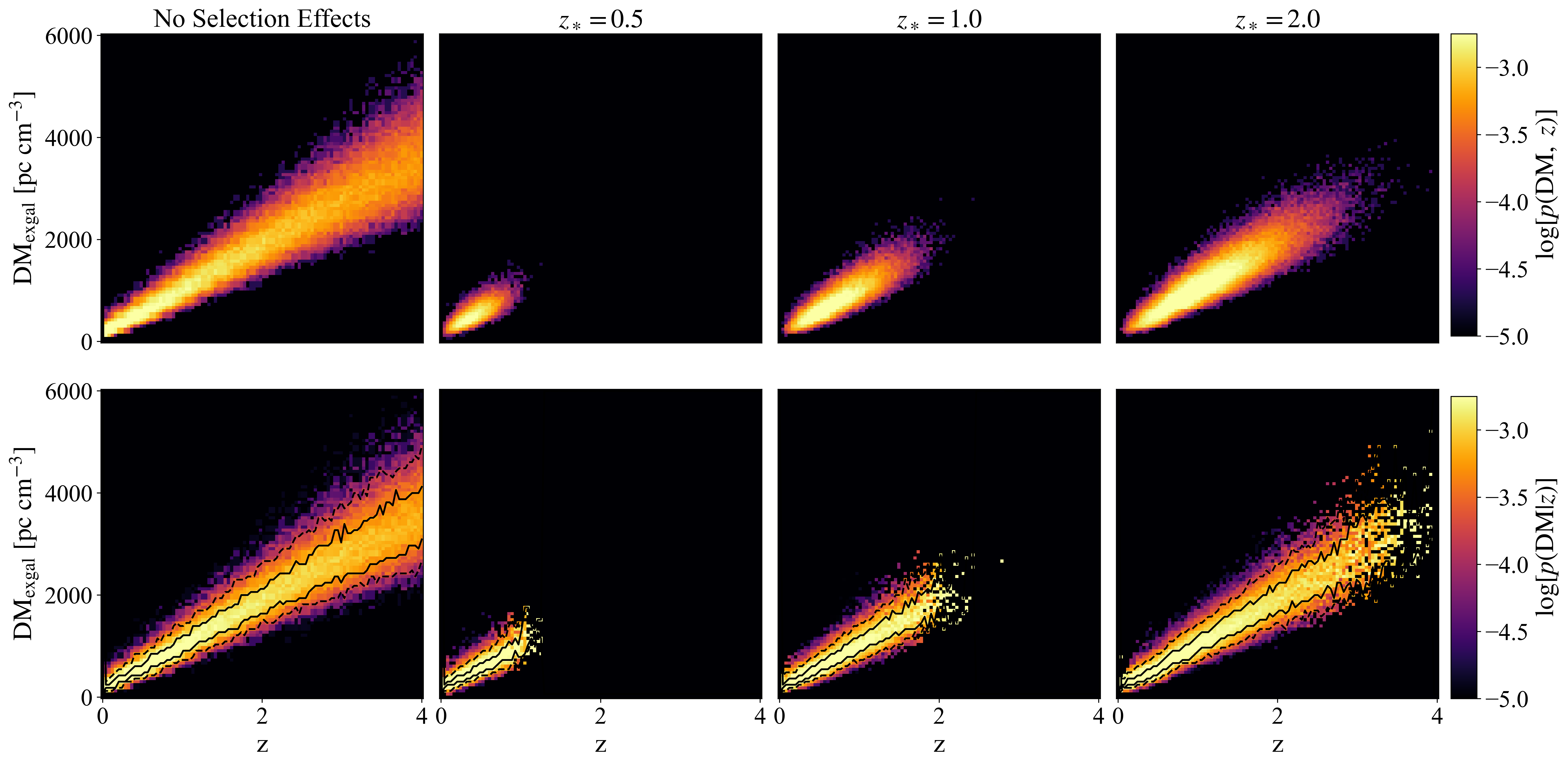}}
\fbox{\includegraphics[width=\textwidth]{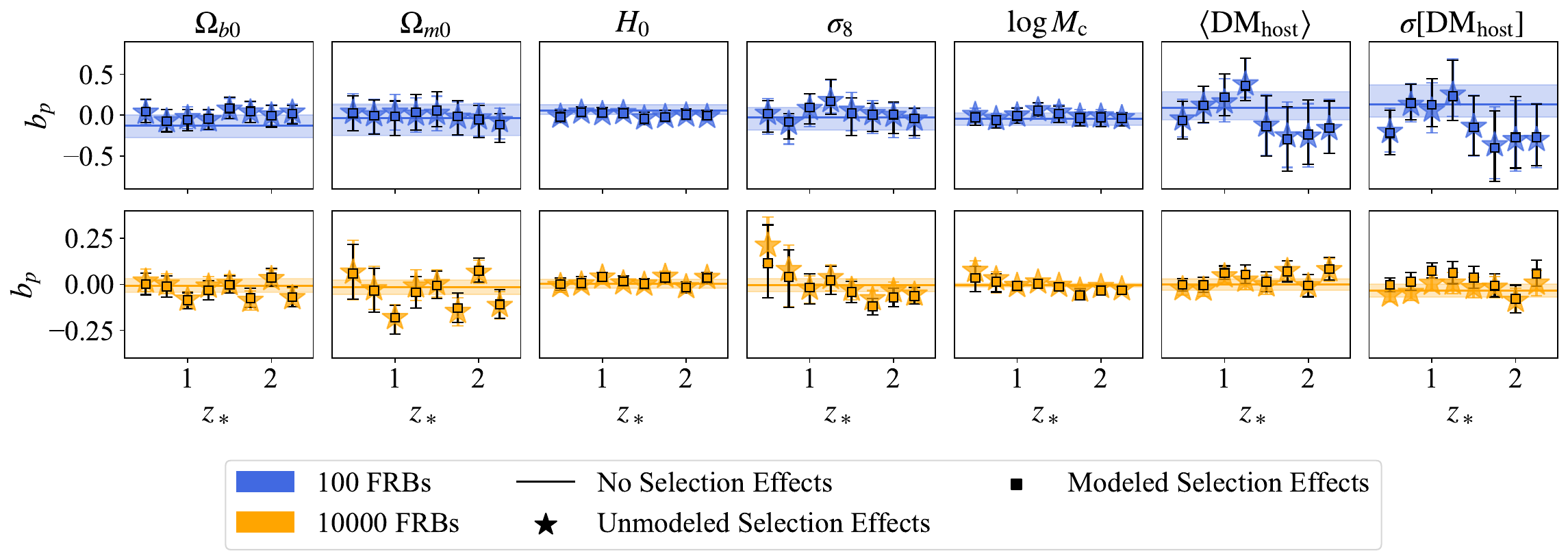}}
\caption{Quantifying the impact of population redshift evolution and luminosity function on cosmological parameter inference conducted using $p(\mathrm{DM}_\mathrm{exgal}|z)$. Top panel illustrates the impact of FRB redshift distribution (controlled by $z_\ast$) on $p(\mathrm{DM}_\mathrm{cosmic}, z)$ (first row) and $p(\mathrm{DM}_\mathrm{cosmic} | z)$ (second row) distributions. The black lines indicate the 68\% and 95\% confidence intervals of $p(\mathrm{DM}_\mathrm{exgal}|z)$. Bottom panel quantifies the bias in inferred parameters, $b_p = (p-p^\mathrm{True})/p^\mathrm{True}$, for various FRB redshift distributions ($z_\ast$). The solid line indicates the inferred bias in no selection effects scenario, which is consistent with zero within $1\sigma$ (shaded region). The bias in parameters when inference is conducted using $p(\mathrm{DM}_\mathrm{exgal}|z)$ (without modeling any selection effects) is shown as stars and when the inference is conducted using $p(\mathrm{DM}_\mathrm{FRB}, z)$ (by modeling selection effects) is shown as squares. For a sample of $10^2$ FRBs (blue), without modeling selection effects, the bias in parameters is consistent with zero within $1\sigma$. As the constraining power increases with $10^4$ FRBs (yellow), the bias in parameters still remains $\lesssim 1\sigma$ when selection effects are not accounted. The bias remains $\lesssim 1\sigma$ after modeling the selection effects.}
\label{fig:pDMz_bias_zdist}
\end{figure*}

We begin by considering the idealized scenario in which the distribution of observed FRBs in the DM$_\mathrm{FRB}$--$z$ plane is unaffected by any selection effects. This setup allows us to benchmark the intrinsic constraining power of FRB samples. The upper-left panel of Figure~\ref{fig:pDMz_bias_zdist} illustrates the distribution of $10^4$ FRBs in the DM$_\mathrm{cosmic}$--$z$ plane. The background color map in the top panel represents the joint probability density $p(\mathrm{DM}_\mathrm{cosmic}, z)$, while that in the bottom panel corresponds to the conditional probability $p(\mathrm{DM}_\mathrm{cosmic}\,|\,z)$. For reference, in the bottom panel, we also show the $1\sigma$ and $2\sigma$ bands. As anticipated, the joint probability density $p(\mathrm{DM}_\mathrm{cosmic}, z)$ decreases along the median DM$_\mathrm{cosmic}$--$z$ relation with increasing redshift. In contrast, the conditional distribution $p(\mathrm{DM}_\mathrm{cosmic}\,|\,z)$ continues to peak around the median of the DM$_\mathrm{cosmic}$--$z$ relation.

Under this idealized scenario, we perform likelihood-based inference using the conditional distribution $p(\mathrm{DM}_\mathrm{exgal}\,|\,z)$ for FRB samples comprising $10^2$ and $10^4$ sources. The inferred fractional bias in each parameter, defined as $b_p = (p - p^\mathrm{true})/p^\mathrm{true}$, is shown in the bottom panels of Figure~\ref{fig:pDMz_bias_zdist} as solid lines, with the shaded regions indicating the $1\sigma$ confidence intervals. As expected, all cosmological, astrophysical feedback, and host-galaxy DM distribution parameters are accurately recovered. The corresponding $1\sigma$ uncertainties for samples of $10^2$ ($10^4$) FRBs are as follows: $\Omega_\mathrm{b0}$ to 25\% (7\%), $\Omega_\mathrm{m0}$ to 36\% (8\%), $H_0$ to 13\% (5\%), $\sigma_{8}$ to 27\% (7\%), $\log M_{\mathrm{c}}$ to 14\% (2\%), $\langle \mathrm{DM}_\mathrm{host} \rangle$ to 37\% (6\%), and $\sigma[\mathrm{DM}_\mathrm{host}]$ to 45\% (7\%). These results quantify the constraining power of FRB samples in the absence of selection effects.

\begin{figure*}[ht!]
\centering
\fbox{\includegraphics[width=\textwidth]{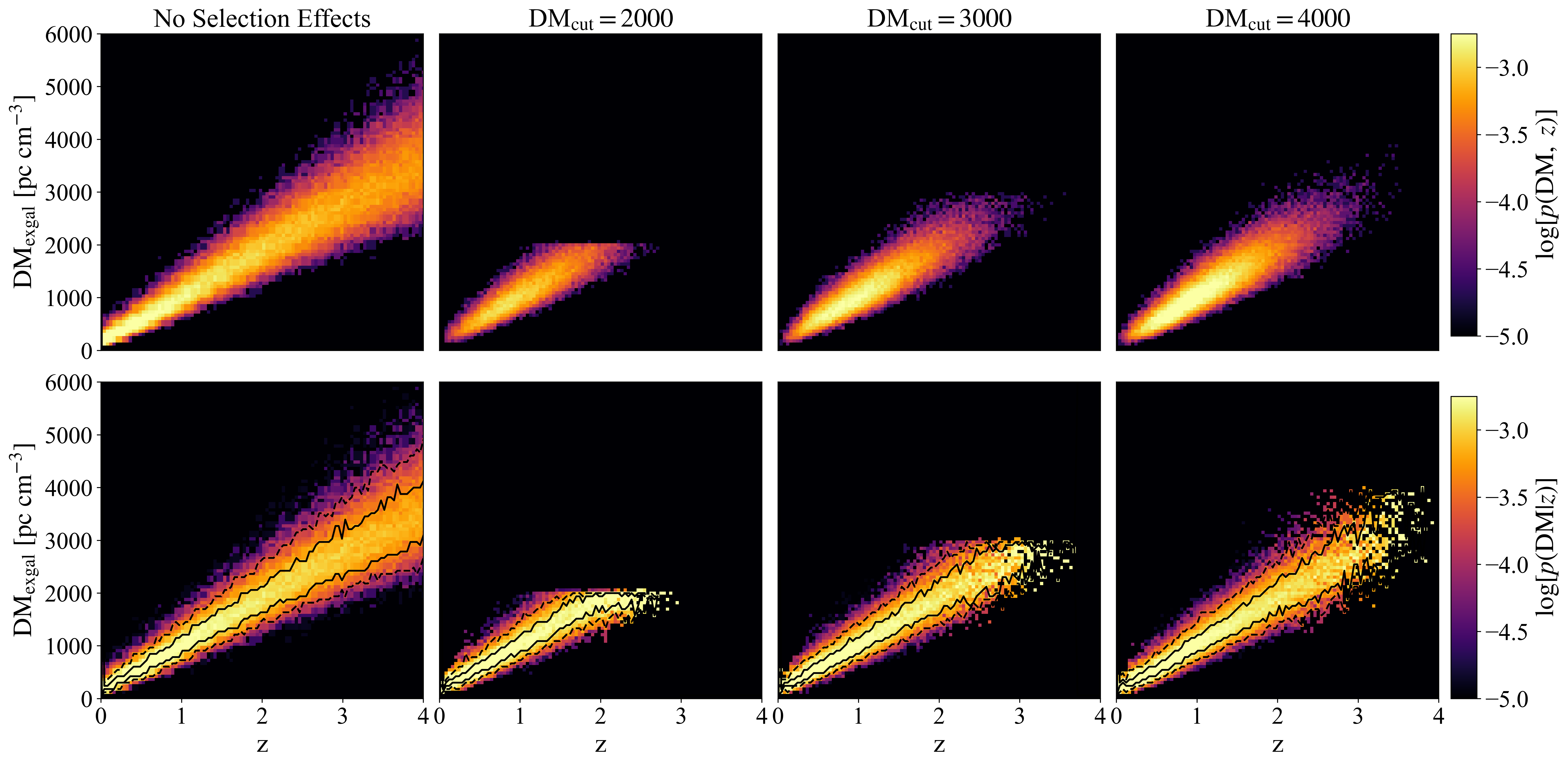}}
\fbox{\includegraphics[width=\textwidth]{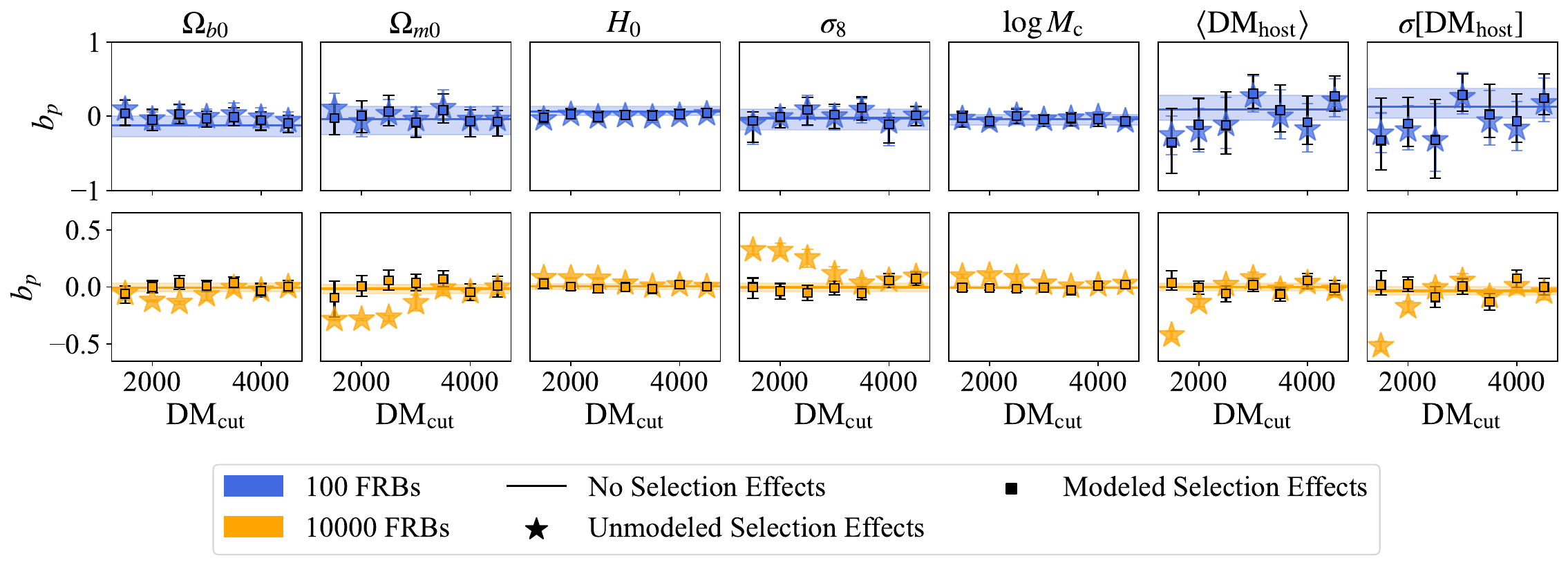}}
\caption{Quantifying the impact of strict DM cuts on cosmological parameter inference conducted using $p(\mathrm{DM}_\mathrm{exgal}|z)$. Top panel illustrates the impact of DM$_\mathrm{cut}$ on $p(\mathrm{DM}_\mathrm{cosmic}, z)$ (first row) and $p(\mathrm{DM}_\mathrm{cosmic} | z)$ (second row) distributions. The black lines indicate the 68\% and 95\% confidence intervals of $p(\mathrm{DM}_\mathrm{exgal}|z)$. Bottom panel quantifies the bias in inferred parameters, $b_p = (p-p^\mathrm{True})/p^\mathrm{True}$, for various DM$_\mathrm{cut}$. The solid line indicates the inferred bias in no selection effects scenario, which is consistent with zero within $1\sigma$ (shaded region). The bias in parameters when inference is conducted using $p(\mathrm{DM}_\mathrm{exgal}|z)$ (without modeling any selection effects) is shown as stars and when the inference is conducted using $p(\mathrm{DM}_\mathrm{FRB}, z)$ (by modeling selection effects) is shown as squares. For a sample of $10^2$ FRBs (blue), without modeling selection effects, the bias in parameters is consistent with zero within $1\sigma$. As the constraining power increases with $10^4$ FRBs (yellow), the bias in parameters can be larger than $3\sigma$ if selection effects are not accounted. Modeling selection effects removes this bias.}
\label{fig:pDMz_bias_zstar2p0_strictDMcut}
\end{figure*}

\begin{figure*}[ht!]
\centering
\fbox{\includegraphics[width=\textwidth]{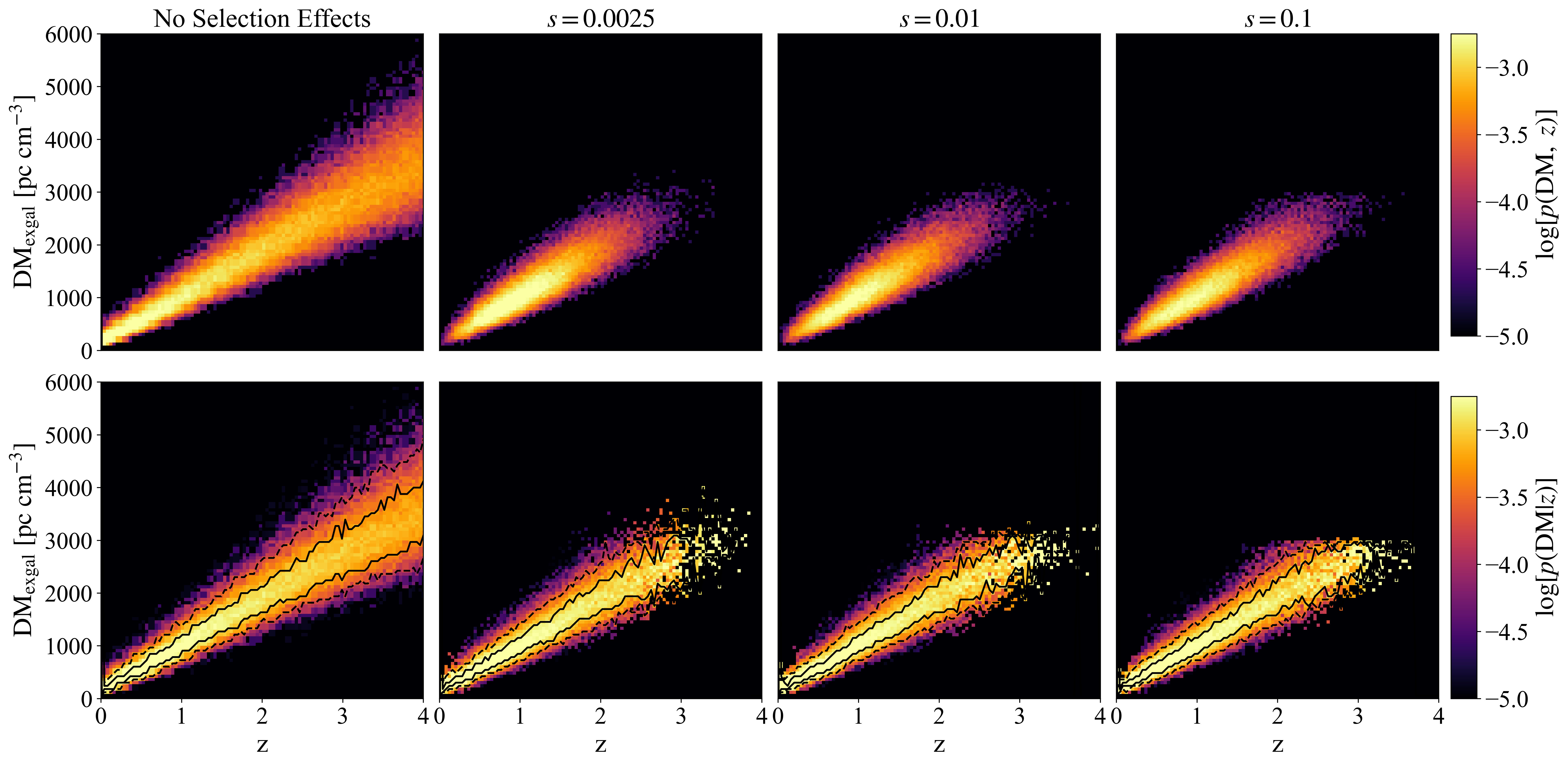}}
\fbox{\includegraphics[width=\textwidth]{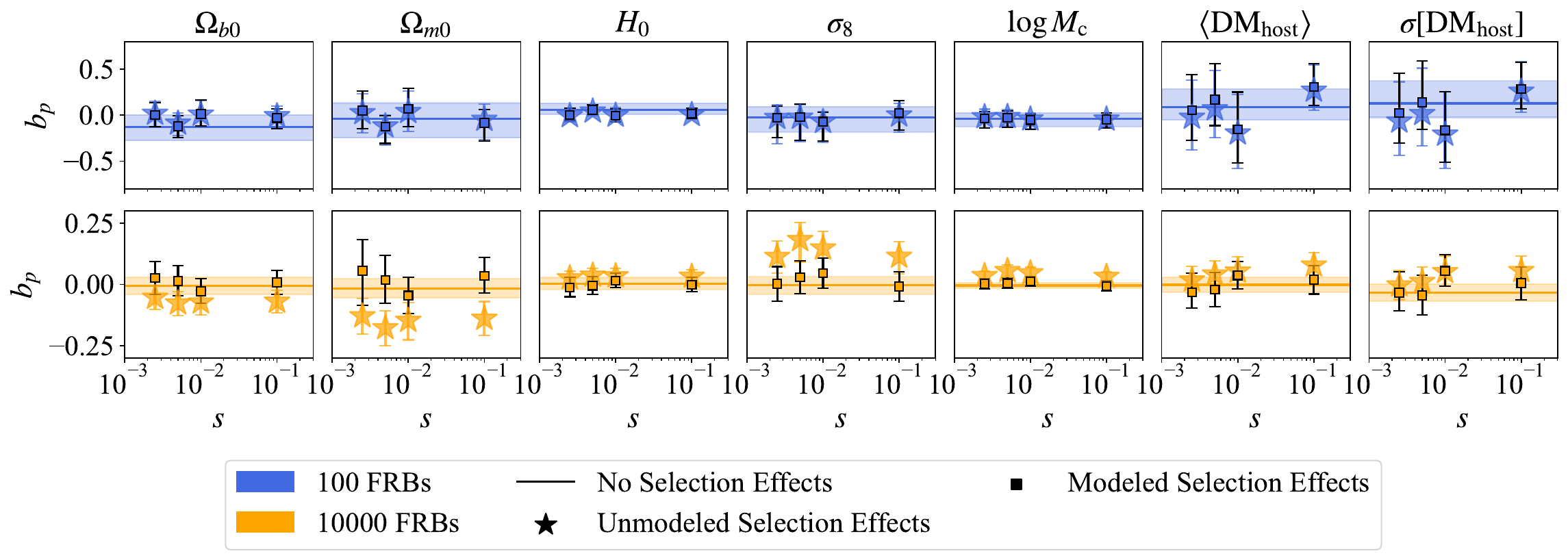}}
\caption{Quantifying the impact of the strength of DM cut on cosmological parameter inference conducted using $p(\mathrm{DM}_\mathrm{exgal}|z)$. Top panel illustrates the impact of DM$_\mathrm{cut}$ strength (denoted by $s$) on $p(\mathrm{DM}_\mathrm{cosmic}, z)$ (first row) and $p(\mathrm{DM}_\mathrm{cosmic} | z)$ (second row) distributions. The black lines indicate the 68\% and 95\% confidence intervals of $p(\mathrm{DM}_\mathrm{exgal}|z)$. Bottom panel quantifies the bias in inferred parameters, $b_p = (p-p^\mathrm{True})/p^\mathrm{True}$, for various DM$_\mathrm{cut}$. The solid line indicates the inferred bias in no selection effects scenario, which is consistent with zero within $1\sigma$ (shaded region). The bias in parameters when inference is conducted using $p(\mathrm{DM}_\mathrm{exgal}|z)$ (without modeling any selection effects) is shown as stars and when the inference is conducted using $p(\mathrm{DM}_\mathrm{FRB}, z)$ (by modeling selection effects) is shown as squares. For a sample of $10^2$ FRBs (blue), without modeling selection effects, the bias in parameters is consistent with zero within $1\sigma$. As the constraining power increases with $10^4$ FRBs (yellow), the bias in parameters can be larger than $1\sigma$ if selection effects are not accounted. Modeling selection effects removes this bias and the loss in constraining power from additional parameters is negligible.}
\label{fig:pDMz_bias_zstar2p0_DMcut}
\end{figure*}

\subsection{FRB source redshift distribution} \label{subsec:imposing_redshift_distribution}

In the first experiment, we incorporate the effects of FRB population redshift evolution and luminosity function, thereby introducing a realistic redshift distribution in the two-dimensional DM$_\mathrm{FRB}$--$z$ plane. Specifically, we adopt a power-law FRB luminosity function with index $\gamma = -0.5$, a minimum detectable redshift of $z_\mathrm{min} = 0$, a redshift evolution of the volumetric FRB source density proportional to the cosmic star formation rate density ($n = 1$) and a spectral index of $\alpha = -0.5$. To modulate the redshift distribution, we vary the detection horizon corresponding to the upper cut-off energy of the luminosity function, parameterized by $z_\ast$. For a given intrinsic (unknown) $E_\ast$ of the FRB population, the resulting redshift distribution extends to higher redshifts for instruments with greater sensitivity (i.e., lower effective fluence thresholds $F_\mathrm{th,eff}$), corresponding to larger values of $z_\ast$.

We illustrate the influence of $z_\ast$ on the DM$_\mathrm{FRB}$--$z$ distribution of FRB sources in Figure~\ref{fig:pDMz_bias_zdist}. The top panels depict the variation of the joint probability density $p(\mathrm{DM}_\mathrm{cosmic}, z)$ with $z_\ast$, while the bottom panels show the corresponding conditional probability $p(\mathrm{DM}_\mathrm{cosmic}\,|\,z)$. Compared to the idealized case without selection effects, the joint probability density $p(\mathrm{DM}_\mathrm{cosmic}, z)$ is strongly modulated by the imposed FRB redshift distribution, whereas the conditional probability $p(\mathrm{DM}_\mathrm{cosmic}\,|\,z)$ remains largely unaffected across the accessible redshift range. This qualitative behavior suggests that inferences based on $p(\mathrm{DM}_\mathrm{cosmic}\,|\,z)$ are expected to be largely insensitive to the underlying FRB redshift distribution. In the following, we quantify this observed qualitative effect.

We perform likelihood-based inference using the conditional distribution $p(\mathrm{DM}_\mathrm{cosmic}\,|\,z)$. For fixed sample sizes of $10^2$ and $10^4$ FRBs, we find that the constraining power on $\sigma_8$ and $\log M_{\mathrm{c}}$ strengthens with increasing $z_\ast$, as the cosmic component constitutes a larger fraction of the total dispersion measure at higher redshifts. For samples of $10^2$ ($10^4$) FRBs, the $1\sigma$ uncertainties on $\sigma_8$ improve from 28\% (9\%) to 26\% (8\%), and on $\log M_{\mathrm{c}}$ from 15\% (3\%) to 14\% (2\%) as $z_\ast$ increases from 0.5 to 2.25. This underscores the critical role of high-redshift FRBs in constraining $\sigma_8$ and probing astrophysical feedback processes.  

Conversely, the constraining power on the host-galaxy DM distribution parameters decreases as $z_\ast$ increases, since at lower redshifts the host contribution comprises a larger fraction of the total extragalactic DM, whereas at higher redshifts it diminishes approximately as $(1+z)^{-1}$. For samples of $10^2$ ($10^4$) FRBs, the $1\sigma$ uncertainties on $\langle \mathrm{DM}_\mathrm{host} \rangle$ improve from 32\% (7\%) to 28\% (6\%), and on $\sigma[\mathrm{DM}_\mathrm{host}]$ from 33\% (7\%) to 30\% (6\%) as $z_\ast$ decreases from 2.25 to 0.5. This highlights the complementary importance of low-redshift FRBs in constraining the host DM distribution.

We show the bias in inferred parameters using the conditional distribution $p(\mathrm{DM}_\mathrm{cosmic}\,|\,z)$ as stars in the bottom panels of Figure~\ref{fig:pDMz_bias_zdist}. As expected, for a small sample of $10^2$ FRBs, the impact of the redshift distribution on the inferred parameters is negligible, with the resulting bias remaining below $\sim 0.8\sigma$. However, for a larger sample of $10^4$ FRBs, the bias can increase to $\sim 0.7$--$1.2\sigma$, depending on the redshift distribution. 

This motivates the need to model redshift-dependent selection effects, including FRB population evolution and luminosity functions. In particular, we fit for the power-law index of the FRB luminosity function, $\gamma$, and the detection horizon corresponding to the upper cut-off energy of the luminosity function, $z_\ast$, adopting broad priors as listed in Table~\ref{table:parameters}. The resulting bias in the inferred parameters, after accounting for selection effects in the joint distribution $p(\mathrm{DM}_\mathrm{cosmic}, z)$ inference, is shown as squares in the bottom panels of Figure~\ref{fig:pDMz_bias_zdist}. As expected, this procedure mitigates biases in most parameters. A few exceptions, notably $\Omega_\mathrm{m0}$ and $\sigma_8$, remain apparent. Several parameters, including $\sigma_8$ and $\log M_{\mathrm{c}}$, exhibit strong degeneracies that cannot be resolved with a low-redshift FRB sample. Such degeneracies are not fully captured in the marginalized bias estimates presented in Figure~\ref{fig:pDMz_bias_zdist}. To assess whether these residual biases are intrinsic, we fix all parameters except $\Omega_\mathrm{m0}$ and $\sigma_8$ and attempt to recover them. We find that these parameters are recovered without bias, indicating that the observed deviations primarily arise from correlations between parameters.

\subsection{Sharp DM cut-off in FRB search} \label{subsec:sharp_DMcut}

The next set of experiments involve imposing a sharp dispersion measure cut-off in the FRB search. For these analyses, we adopt a detection horizon corresponding to the upper cut-off energy of the luminosity function at $z_\ast = 2$, ensuring a sufficiently large redshift lever arm to explore the impact of DM cuts over an extended range. Figure~\ref{fig:pDMz_bias_zstar2p0_strictDMcut} presents the joint probability distribution $p(\mathrm{DM}_\mathrm{cosmic}, z)$ and the conditional distribution $p(\mathrm{DM}_\mathrm{cosmic}\,|\,z)$ for $10^4$ simulated FRBs, with $\mathrm{DM}_\mathrm{cut}$ varied between 2000 and 4000~pc\,cm$^{-3}$. As expected, when compared to the no-selection scenario, the joint distribution $p(\mathrm{DM}_\mathrm{cosmic}, z)$ is significantly distorted across all values of $\mathrm{DM}_\mathrm{cut}$. In contrast, the conditional distribution $p(\mathrm{DM}_\mathrm{cosmic}\,|\,z)$ remains largely unaffected within the accessible redshift range. 

Following the same procedures outlined in Section~\ref{subsec:imposing_redshift_distribution}, we begin our analysis by performing likelihood inference using $p(\mathrm{DM}_\mathrm{cosmic}\,|\,z)$. The resulting biases in the inferred cosmological, astrophysical feedback, and host-galaxy DM distribution parameters are shown as stars in the bottom panels of Figure~\ref{fig:pDMz_bias_zstar2p0_strictDMcut}. As expected, for a small sample of $10^2$ FRBs, the resulting parameter biases are $\lesssim 0.5\sigma$. However, as the sample size increases to $10^4$ FRBs, the biases can be $\gtrsim 3\sigma$. Such significant biases in the inferred parameters highlight the necessity of modeling the DM selection function, particularly in the large-sample regime.

Next, we perform likelihood inference using the joint probability distribution $p(\mathrm{DM}_\mathrm{FRB}, z)$, where, in addition to the model parameters introduced in Section~\ref{subsec:imposing_redshift_distribution}, we also fit for the DM selection function, specifically the $\mathrm{DM}_\mathrm{cut}$ with the wide prior listed in Table~\ref{table:parameters}. The resulting biases in all inferred parameters for different $\mathrm{DM}_\mathrm{cut}$ values are shown as squares in the bottom panels of Figure~\ref{fig:pDMz_bias_zstar2p0_strictDMcut}. As expected, when incorporating this extended model, the biases in parameter estimates reduce to $\lesssim 0.6\sigma$.

\subsection{Varied strength of DM cut-off in FRB search} \label{subsec:vary_DMcut}

In this final experiment, we examine the impact of varying the strength of the $\mathrm{DM}_\mathrm{cut}$. For the case with $\mathrm{DM}_\mathrm{cut} = 3000~\mathrm{pc\,cm^{-3}}$, we vary the strength parameter $s \in (10^{-3}, 10^{-1})$. The corresponding effects on the joint probability distribution $p(\mathrm{DM}_\mathrm{cosmic}, z)$ and the conditional distribution $p(\mathrm{DM}_\mathrm{cosmic}\,|\,z)$ for $10^4$ simulated FRBs are illustrated in the top panels of Figure~\ref{fig:pDMz_bias_zstar2p0_DMcut}. Consistent with findings in former sections, the impact of DM selection effects on the conditional distribution within the accessible redshift range remains negligible. When performing likelihood inference using conditional probability distribution $p(\mathrm{DM}_\mathrm{cosmic}\,|\,z)$  with a small sample of $10^2$ FRBs, the bias in all inferred parameters remains $\lesssim 0.6\sigma$ (shown as blue stars in the bottom panels of Figure~\ref{fig:pDMz_bias_zstar2p0_DMcut}). However, for a larger sample of $10^4$ FRBs, the systematic bias can increase up to $1.3\sigma$ (shown as yellow stars in the bottom panels of Figure~\ref{fig:pDMz_bias_zstar2p0_DMcut}). Incorporating a model for the DM selection function with varying cut-off strengths into the inference based on $p(\mathrm{DM}_\mathrm{FRB}, z)$, and adopting broad priors as listed in Table~\ref{table:parameters}, effectively mitigates these biases, reducing them to $\lesssim 0.4\sigma$.

\subsection{Limitations of our Parameterization}\label{subsec:limitations}

So far, we have shown that the impact of selection effects on cosmological inference using FRB DM--$z$ relation is not significant for existing FRB sample sizes and introducing additional model complexity is not essential at this stage. Below, we discuss the limitations of our parameterization of selection effects and/or possible extensions to the model used which may be useful for next generation FRB surveys (and/or other science cases).

\begin{itemize}[leftmargin=*]

    \item In our work, we assume that the detection probability of an FRB is separable in DM and redshift space as
    \begin{equation}
        p_\mathrm{detect}(\mathrm{DM}_\mathrm{FRB}, z) = S(\mathrm{DM}_\mathrm{FRB}) \cdot p_\mathrm{detect}(z).
    \label{eqn:p_detect_coupled}
    \end{equation}
    However, technically, the detection probability of a given DM and redshift is coupled because the effective burst width, and hence the effective fluence threshold (which defines redshift sensitivity in our parameterization), depends on observed DM through smearing (see equation~\ref{eqn:effective_burst_width}). As the redshift of the burst increases, the DM smearing of the burst increases, consequently increasing the effective fluence threshold of the instrument and reducing the sensitivity to high-redshift FRBs. On the contrary, the $(1+z)^{-3}$ suppression of scattering timescales are expected to increase the FRB detection rates by $\sim 10$\% at high-redshift~\citep{2025arXiv251005654J}. This interplay of DM smearing and scattering timescales may alter the observed two-dimensional DM--$z$ distribution, impacting accuracy of cosmological parameters in the era of large number statistics. 

    \item Since the objective of this work was cosmological inference (and not luminosity function estimation), we chose not to include the signal-to-noise ratio (SNR) dimension in our model. The SNR of detected bursts primarily constrains the effective fluence threshold, which can be used to convert the measured detection horizon $z_\ast$ to an upper cut-off energy for the luminosity function. For applications focused on inferring the FRB luminosity function, our framework can be readily extended to include the SNR $s$ of bursts via
    \begin{equation}
        p(s \,|\, \mathrm{DM}_\mathrm{FRB}, z) = \frac{p(E = s\,E_\mathrm{th}(z))}{p(E \geq E_\mathrm{th}(z))}.
    \end{equation}

    \item Whether and how DM$_\mathrm{host}$ redshift evolution should be modeled in DM--$z$ analysis is not known; significant evolution is expected if FRBs trace star-formation rate~\citep{2024ApJ...972L..26O}. If DM$_\mathrm{host}$ evolves with redshift as $\propto (1+z)^1$, then the DM$_\mathrm{exgal}$--$z$ analysis for a sample of $10^4$ FRBs can be significantly biased if this evolution is ignored. Our simulations suggest that the inferred cosmological parameters can be biased by $3-6\sigma$, thus necessitating better understanding and modeling of DM$_\mathrm{host}$ redshift evolution.

    \item If the volumetric FRB rate is assumed to evolve proportionally to the cosmic star-formation history (SFH) without incorporating an explicit delay-time distribution (DTD), the inferred source redshift distribution will be incorrectly modeled, introducing bias in DM--$z$ based cosmological analyses. Physically, if a significant fraction of FRB progenitors arise from channels with non-negligible delay times --- such as compact object mergers or delayed core-collapse supernovae --- then the true redshift evolution of the population should be given by a convolution of the SFH with the modeled DTD. However, quantifying the impact of FRB DTD on cosmological inferences is beyond the scope of this work.

\end{itemize}

\section{Conclusion} \label{sec:conclusion}

In this work, we have systematically quantified the impact of FRB population characteristics and instrument selection effects on cosmological inference conducted using the conditional probability distribution $p(\mathrm{DM}_\mathrm{exgal} | z)$, without modeling any observational biases. We developed a forward-modeling framework to emulate FRB populations and performed likelihood-based inference under progressively realistic observational scenarios. We now provide a summary of the principal outcomes of our work.

\begin{itemize}[leftmargin=*]

    \item For rapid and accurate likelihood evaluations in MCMC analysis without repeated halo model calculations, we built a neural network emulator for the variance in DM$_\mathrm{cosmic}$ as a function of four cosmological parameters and one astrophysical feedback parameter $\log M_c$ (see Section~\ref{sec:building_emulator}). This model, trained on $5\times 10^4$ baryonification halo model simulations, exhibits prediction errors of $\leq 4\%$ upto $z = 4$ on the validation dataset, thus establishing its accuracy (see Figure~\ref{fig:emulator}).
    
    \item Modeling of the observational biases can be significantly simplified by assuming that the detection probability of an FRB event is separable in DM and $z$, with a constant $F_\mathrm{th,eff}$ marginalized over burst width distributions and telescope beam response (see Section~\ref{sec:model}). This serves as a reasonable first-order approximation--eliminating the need to explicitly model uncertain redshift-dependent intrinsic and scattering width distributions or beam patterns. While the framework can be extended to include more complex, FRBs- and instrument-specific effects, as demonstrated in this work, such additional complexity may not be essential for current FRB samples.
    
    \item In the idealized scenario without selection effects, a sample of $10^4$ FRBs offers sub-10\% precision on $\Omega_\mathrm{b0}$, $\Omega_\mathrm{m0}$, $H_0$, $\sigma_8$, and $\log M_\mathrm{c}$. Introducing redshift-dependent FRB population evolution and luminosity functions, together with instrument DM selection function, impacts the joint distribution $p(\mathrm{DM}_\mathrm{cosmic}, z)$ but leaves the conditional distribution $p(\mathrm{DM}_\mathrm{cosmic}\,|\,z)$ largely unaffected (see Section~\ref{sec:quantify_biases}). Consequently, inference based on $p(\mathrm{DM}_\mathrm{cosmic}\,|\,z)$ remains robust to such selection effects with induced biases up to $\sim 0.8\sigma$ for $10^2$ FRBs and $\gtrsim 3\sigma$ for $10^4$ FRBs. Modeling the redshift-dependent FRB luminosity function and selection function within the inference framework effectively mitigates these biases, reducing them to $\lesssim 1\sigma$.
    
    \item We find that high-redshift FRBs provide stronger constraints on cosmological and feedback parameters ($\sigma_8$, $\log M_\mathrm{c}$), while low-redshift FRBs are crucial for constraining the host DM distribution. A balanced FRB sample spanning a broad redshift range is therefore essential for breaking parameter degeneracies.
\end{itemize}

Overall, our results highlight that while redshift- and DM-dependent selection effects do not significantly bias cosmological parameter inference for small FRB samples, accurately modeling these effects may become important as upcoming surveys are expected to localize thousands of  FRBs.

\begin{acknowledgments}
K.S. thanks Clancy James and Xavier Prochaska for insightful conversations at FRB 2024 meeting that have helped shape this paper.
\end{acknowledgments}

\bibliography{manuscript}{}
\bibliographystyle{aasjournal}

\end{document}